\documentclass[11pt]{article}
\usepackage[colorlinks=true]{hyperref}
\usepackage[margin=1in]{geometry}
\usepackage{setspace}
\usepackage{epsfig,amssymb,latexsym,bm}
\usepackage{amsfonts,psfrag,amsmath,mathrsfs,amsthm,color,empheq}
\usepackage{graphicx}

\usepackage{enumerate}
\usepackage{float}
\usepackage[nocompress]{cite}

\usepackage{mathtools,soul}
\usepackage[displaymath, mathlines]{lineno}
\usepackage{tikz}
\usetikzlibrary{arrows.meta,positioning}

\newcommand{\lastdate}{April 30, 2020}

\def\a{\alpha}
\def\b{\beta}
\def\d{\delta}

\def\g{\gamma}

\graphicspath{{./Figures/}}
\usepackage{epstopdf}
\title{Optimal allocation of face masks during the COVID-19 pandemic: a case study of the first epidemic wave in the United States}
\author{Jun Liu\thanks{
Department of Mathematics and Statistics, Southern Illinois University Edwardsville,  Edwardsville, IL 62026, USA. Email: juliu@siue.edu} ~and~ Xiang-Sheng Wang\thanks{
Department of Mathematics, University of Louisiana at Lafayette, Lafayette, LA 70503, USA. Email: xiangsheng.wang@louisiana.edu}}

\begin{document}

\maketitle

\begin{abstract}
In this paper, we propose a two-group SIR epidemic model to simulate the outcome of stay-at-home policy and wearing face masks during the first COVID-19 epidemic wave in the United States. Based on our proposed model, we further use the optimal control approach (with the objective of minimizing total deaths) to find the optimal dynamical distribution of face masks between the healthcare workers and the general public. It is not surprising that all the face masks should be solely reserved for the healthcare workers if the supply is short. However, when the supply is indeed sufficient, our numerical study indicates that the general public should share a large portion of face masks at the beginning of an epidemic wave to dramatically reduce the death toll.  This interesting result partially contradicts with the guideline advised by the US Surgeon General and the Centers for Disease Control and Prevention (CDC) in March 2020. The optimality of this sounding CDC guideline highly depends on the supply level of face masks that changes frequently, and hence it should be adjusted according to the supply of face masks.
\end{abstract}


\section{Introduction}

Ever since the first reported case on December 31, 2019, in Wuhan, China \cite{FirstCase}, the ongoing outbreak of COVID-19 has been spreading around the whole world.
During the first two months of 2020, the pandemic was mainly restricted in mainland China, especially within the Hubei province. On March 1, 2010, there were 87,137 reported cases in the world \cite{SituationReports}, more than 90\% of which were from China, and 67,103 cases were reported in Hubei province \cite{China}.
In March 2020, the epidemic wave in China was approaching the end, while the reported case numbers in Europe and North America grew rapidly. Even though the first case was reported on January 22, 2020 \cite{Cases-in-US}, the United States did not take very serious or effective actions until March 2020 \cite{Emergency}. Due to different cultures and medical systems, the control strategies vary from one country to another. For instance, wearing face masks for everyone in public areas is ubiquitous and even enforced by compulsory policies in China, Japan and South Korea \cite{Feng2020LancetRM}.
However, in March 2020, the US Surgeon General advised the public general not to buy face masks \cite{StopBuying}, and the Centers for Disease Control and Prevention (CDC) commented that face masks won't protect the healthy people from getting the SARS-CoV-2 \cite{MaskNoProtection}. Nevertheless, according to the study in \cite{vD2020NEJM,Howard2020}, the aerosol transmission of SARS-CoV-2 is possible because the virus can last in aerosols for hours and remain viable and infectious. Moreover, there are plenty of scientific evidence showing that a person not developing any symptom may still be infectious \cite{Chan20Lancet}. Thus, wearing face masks in public is an effective way to reduce the spreading of COVID-19, and more importantly to prevent those asymptomatic carriers \cite{Arons2020} from infecting others.

In several recent papers \cite{eikenberry2020mask,ngonghala2020,Worby2020}, some dedicated compartmental models are developed for assessing the community-wide impact of massively using face masks by the general public, which shows its high efficacy in curtailing community transmission and reducing the burden of the pandemic. However, to the best of our knowledge, there is no related discussion about how to optimize the allocation of face masks between healthcare workers (HCW) and the general public. In April 2020, the CDC started to recommend wearing cloth face coverings for the general public (GP). Due to initial shortage in supply, the CDC advised to save face masks for healthcare workers \cite{Protect}. It poses a natural question to ask: whether one should distribute a portion or none of face masks to the general public along the epidemic outbreak? In other words, should the general public continue reserves all the face masks for the HCW only?
We are particularly interested in the group of HCW since higher nurse-to-patient staffing ratios can result in healthier patients
and hence protecting the health of the limited number of HCW is paramount  \cite{Black2020,Miller2020}. In general, it may takes several years to train a HCW who are qualified to do the professional job.
The objective of this work is to design an epidemic model with face masks and stay-at-home factors to find the optimal dynamical distribution of face masks among the healthcare workers and the general public, with the purpose of minimizing the total number of deaths.

This paper is organized as follows.
In the next section, an optimal control model based on a new two-group SIR model is introduced,
where the effects of wearing face masks and stay-at-home policy are considered.
The justification and estimation of the relevant parameters
of our proposed model is discussed in Section 3.
Section 4 presents the numerical results, where three different scenarios are demonstrated.
Finally, some conclusion and discussion are given in Section 5.
\section{A Two-Group SIR Model}
\begin{figure}[H]
	\begin{center}
		\begin{tikzpicture}[node distance=4cm, auto,
		>=Latex, thick,
		every node/.append style={align=center},
		int/.style={draw, minimum size=1cm}]
		
		\node [int] (S1)             {$S_1$};
		\node [int, right=of S1] (I1) {$I_1$};
		\node [int, right=of I1] (R1) {$R_1$};
		
		\node [circle,int, below=1.5cm of R1] (D) {Death};
		
		\node [int, below=of S1] (S2)             {$S_2$};
		\node [int, right=of S2] (I2) {$I_2$};
		\node [int, right=of I2] (R2) {$R_2$};
		\node [int, below left=2cm of S1] (Out) {Isolated};
		
		\path[->] (S1) edge node {$\beta f_1S_1(f_1I_1+f_1I_2)$} (I1)
		(S1) edge node {$qS_1$} (Out)
		(S2) edge node[below] {$\beta f_2S_2(f_1I_1+f_1I_2)$} (I2)
		(I1) edge node {$\gamma I_1$} (R1)
		(I1) edge node {$\delta I_1$} (D)
		(I2) edge node[below] {$\gamma I_2$} (R2)
		(I2) edge node {$\delta I_2$} (D)
		(I1) edge[dashed,out=-150, in=-30] node[above] {$\beta f_1S_1(f_1I_1)$} (S1)
		(I1) edge[dashed,out=-140, in=90] node[right] {$\beta f_2S_2(f_1I_1)$} (S2)
		(I2) edge[dashed,out=140, in=-90] node[right] {$\beta f_1S_1(f_1I_2)$} (S1)
		(I2) edge[dashed,out=150, in=30] node {$\beta f_2S_2(f_1I_2)$} (S2)
		;
		\end{tikzpicture}
	\end{center}
	\caption{A  transfer  diagram  of our proposed two-group SIR model (\ref{S1}-\ref{R2}). Here the solid arrow represents a flux of individuals, while the dashed arrows represent the influence of infectious groups.}
	\label{Model}
\end{figure}
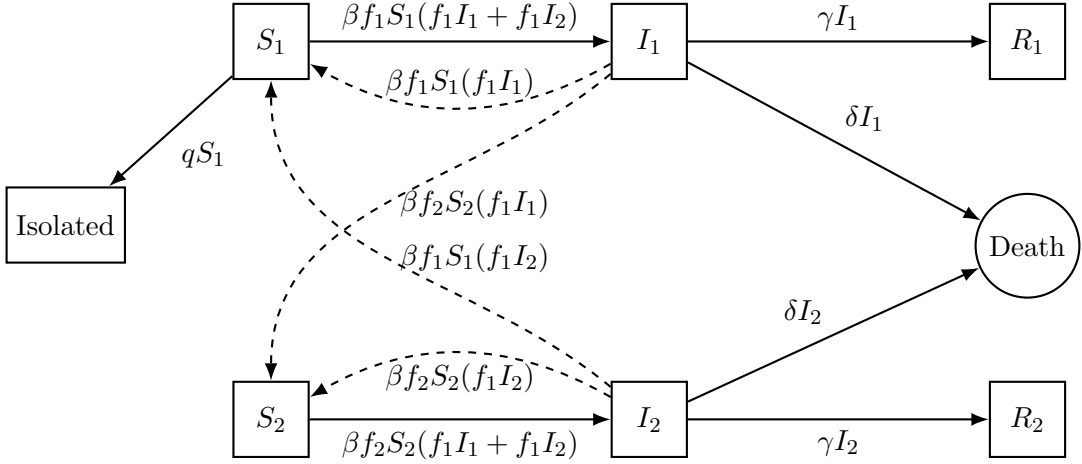

Our model is based on the standard SIR (susceptible-infected-recovered) model proposed in \cite{Kermack1927}.
To distinguish between two groups: the general public (GP) and healthcare workers (HCW), we introduce six compartments $S_1,I_1,R_1,S_2,I_2,R_2$ for the SIR classifications in the two groups. The subscripts $1$ and $2$ refer to the GP and the HCW, respectively.
We do not include the asymptomatic (or latent) compartment because we assume that all infected individuals are infectious even though they do not develop any symptoms \cite{Chan20Lancet}.
Let $K_1(t)\ge 0$ and $K_2(t)\ge 0$ be the numbers of face masks distributed in the two groups over the given time period $[0,T]$. The average number of available face masks for these two groups are $\rho_1=K_1/(S_1+I_1+I_2)$ and $\rho_2=K_2/(S_2+R_2)$. Here, we assume that the recovered GP ($R_1$) do not need face masks but the recovered HCW ($R_2$) will return their job and still need to wear face masks. Moreover, the infected HCW ($I_2$) are released from their duty of healthcare and thus considered as a part of GP.
Now, we can set the transmission rates for the GP and the HCW to be $\b f_1(\rho_1)$ and $\b f_2(\rho_2)$, where $\b>0$ is the intrinsic transmission rate of SARS-CoV-2, and $f_1$ and $f_2$ are positive and decreasing functions. For simplicity, we choose
\begin{equation}\label{f12}
  f_1(\rho_1)={1+\a\rho_1\over1+\rho_1},~~f_2(\rho_2)={r+\a\rho_2\over1+\rho_2},
\end{equation}
where $1-\a\in(0,1)$ stands for the maximum efficacy of wearing face masks, and $r>1$ indicates the higher risk of infection for the HCW comparing to the GP.
Next, we introduce the ratio of infected individuals over the number of HCW: $\rho=(I_1+I_2)/(S_2+R_2)$ and assume that the recovery rate and death rate depend on $\rho$ according to two given functions $\g(\rho)$ and $\d(\rho)$, respectively. It is reasonable to assume that the higher $\rho$ is, the smaller $\g(\rho)$ and the larger $\d(\rho)$ will be.
In our simulations, we will choose
\begin{equation}\label{gd}
  \g(\rho)={\g_0+\g_\infty\rho\over1+\rho},~~\d(\rho)={\d_0+\d_\infty\rho\over1+\rho},
\end{equation}
where $\g_0>0$ and $\d_0>0$ are the recovery and death rates of an infected individual with sufficient healthcare ($\rho\to0$), and $\g_\infty>0$ and $\d_\infty$ are the rates when the healthcare system is overwhelmed ($\rho\to\infty$). Obviously, $\g_0>\g_\infty$ and $\d_0<\d_\infty$.
To estimate the efficacy of stay-at-home policy, we introduce another parameter $q>0$ to account for the portion of susceptible individuals who limit their activity in an isolated region (staying at home for example) so that they are isolated from the infected group.
Finally, we are ready to state our two-group SIR model which consists of six ordinary differential equation (see Figure \ref{Model} for its transfer diagram):
\begin{align}
  S_1'&=-\b f_1(\rho_1)S_1[f_1(\rho_1)I_1+f_1(\rho_1)I_2]-qS_1,\label{S1}  \\
  S_2'&=-\b f_2(\rho_2)S_2[f_1(\rho_1)I_1+f_1(\rho_1)I_2],\label{S2}\\
  I_1'&=\b f_1(\rho_1)S_1[f_1(\rho_1)I_1+f_1(\rho_1)I_2]-[\g(\rho)+\d(\rho)]I_1,\label{I1}\\
  I_2'&=\b f_2(\rho_2)S_2[f_1(\rho_1)I_1+f_1(\rho_1)I_2]-[\g(\rho)+\d(\rho)]I_2,\label{I2}\\
  R_1'&=\g(\rho)I_1,\label{R1}\\
  R_2'&=\g(\rho)I_2,\label{R2}
\end{align}
where $\rho_1=K_1/(S_1+I_1+I_2)$, $\rho_2=K_2/(S_2+R_2)$, $\rho=(I_1+I_2)/(S_2+R_2)$, the functions $f_1$ and $f_2$ are given in \eqref{f12}, and the functions $\g$ and $\d$ are defined in \eqref{gd}. The prime symbol on the left-hand side of the equations denotes the ordinary derivative in time $t$. Such two-group SIR epidemic models were widely known in literature, see e.g. \cite{Magal2016} and references therein. Note that $f_i(\rho)$ (with $i=1$ or $i=2$) appears twice in the incidence rates. This is because the face masks are more effective in reducing disease transmission if both susceptible and infected individuals are wearing them.
Let $T$ be the considered duration of an outbreak and $K_{\max}$ be the maximum capacity of daily production number of face masks. The objective is to minimize the total number of death
\begin{equation}
\min_{K_1,K_2}\qquad  J(K_1,K_2):=\underbrace{\int_0^T \d(\rho)I_1dt}_{=:J_1}+\underbrace{\int_0^T \d(\rho)I_2dt}_{=:J_2},
\end{equation}
subject to the point-wise control constraints describing the production capacity limits
\begin{equation}
0\le K_1(t),\quad 0\le K_2(t),\quad K_1(t)+K_2(t) \le K_{\max}.
\end{equation}
The above optimization model gives a nonlinear constrained optimal control problem, whose optimal control (may not be unique) can be mathematically characterized via Pontryagin's minimum principle \cite{kirk2012optimal}.
Due to its high non-linearity, we will numerically solve the above optimal control problem using a direct transcription method \cite{betts2010practical,Kelly2017} based on Legendre-Gauss-Radau pseudo-spectral collocation \cite{Garg2010,Patterson2014} and nonlinear programming (NLP) solvers (e.g., IPOPT \cite{Biegler2009} and SNOPT\cite{GilMS05}).
Different from the indirect methods that require to derive necessary optimality conditions, such direct transcription methods are more flexible in treating extra constraints and are also widely supported by  well-developed general optimal control software packages that are ready to be used.
In particular, our following numerical simulations are performed with the free and open-source Imperial College London Optimal Control Software (ICLOCS2) \cite{Nie2018} within MATLAB , where the derivatives are numerically computed by the algorithmic differentiation toolbox Adigator \cite{Weinstein2017}.

\section{Estimation of Parameters}

To estimate the model parameters, we need to fit the reported data on cumulative confirmed case numbers, denoted by $C(t)$.
The change rate of $C(t)$ is the same as the newly infected case number, which according to our model is
\begin{equation}\label{C}
  C'=\b [f_1(\rho_1)S_1+f_2(\rho_2)S_2][f_1(\rho_1)I_1+f_1(\rho_1)I_2].
\end{equation}
The initial numbers of susceptible population and health-care workers are estimated as $S_1(0)=310,000,000$ \cite{Population} and $S_2(0)=16,000,000$ \cite{HCW}.
Throughout this paper, the time unit is in days.
We choose March 13, 2020 as the initial time $t=0$ since an emergency declaration was warranted for the COVID-19 pandemic on that date \cite{Emergency}.
Accordingly, we set $I_1(0)=1,896$ \cite{Cases-in-US}. Since the epidemic wave was starting and stay-at-home policy was adopted following the declaration, we let $I_2(0)=R_1(0)=R_2(0)=0$. Based on the cumulative case numbers reported from February 28, 2020 to March 13, 2020 \cite{Cases-in-US}, we can estimate the intrinsic growth rate as $\b S_1(0)-\g_0-\d_0=0.35$ by fitting a simple exponential growth model
\begin{equation}\label{C1}
 C(t)=C(0)e^{[\b S_1(0)-\g_0-\d_0]t},
\end{equation}
which can be obtained by approximating our model under the assumptions that, before March 13, 2020, stay-at-home policy was not adopted ($q=0$), few people wore face masks ($K_1+K_2\ll S_2\ll S_1$), and the infected population was small ($I_2\ll I_1\ll S_2$). Consequently, we have $S_1(t)\approx S_1(0)$ and $\rho_1,\rho_2,\rho,I_2\approx0$. The equation \eqref{I1} is approximated by a linear equation $I_1'=[\b S_1(0)-\g_0-\d_0]I_1$ with solution $I_1(t)=I_1(0)e^{[\b S_1(0)-\g_0-\d_0]t}$. The equation for the cumulative case numbers $C(t)$ in \eqref{C} can be approximated as $C'(t)=\b S_1(0)I_1(t)=\b S_1(0)I_1(0)e^{[\b S_1(0)-\g_0-\d_0]t}$. This together with the initial condition $C(-\infty)=0$ gives \eqref{C1}.

The basic reproduction number $R_0=\b S_1(0)/(\g_0+\d_0)$ for SARS-CoV-2 was estimated to be about $2.2$ \cite{Li2020NEJM}.
In view of $\b S_1(0)-\g_0-\d_0=0.35$, we obtain from a simple calculation $\g_0+\d_0=0.29$ and $\b S_1(0)=0.64$. To estimate $\g_0$ and $\d_0$, we need to make use of the death rate which varies from $4\%$ to $7.5\%$ \cite{DeathRate}. Here, we choose $\d_0/(\g_0+\d_0)=7\%$. Consequently, we have $\d_0=0.02$ and $\g_0=0.27$. There is no data available for the case when the healthcare system is overwhelmed by too many infected patients, and we simply assume that with a overwhelmed healthcare system the death rate will reach $\d_\infty=0.1$ and the recovery rate reduces to $\g_\infty=0.1$.

The values for $\a,r$ cannot be found in the literature. Here, we set $\a=0.9$ (face masks can at most reduce the transmission rate by  $10\%$), and $r=3$ (the healthcare workers are three times more likely to be infected than the general public). We have used some other values for these parameters, and the results do not vary much. Especially, we numerically find that the optimal distributions of face masks have very similar patterns for any reasonable choices of $\a$ and $r$.

\begin{figure}[H]
	\begin{center}
		\includegraphics[width=0.80\textwidth]{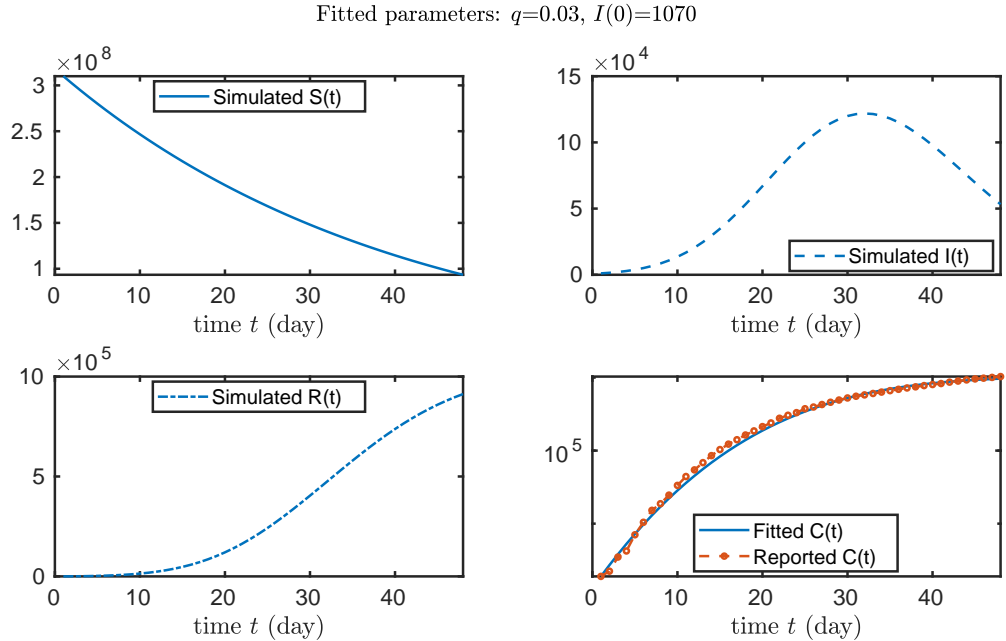}
	\end{center}
	\caption{The system dynamics based on the fitted parameters: $q=0.03, I(0)=1070$.}
	\label{Fitq}
\end{figure}

To estimate the efficacy of stay-at-home parameter $q$, we assume $\rho\approx0$ and ignore the population of healthcare workers ($S_2<<S_1$ and $I_2<<I_1$). For simplicity, we also assume that the majority does not wear face masks ($\rho_1\approx0$) and then fit the confirmed cumulative case numbers reported from March 13, 2020 to \lastdate \cite{Cases-in-US} by the following simplified SIR model:
\begin{align}
  S'&=-\b SI-qS,\\
  I'&=\b SI-(\g_0+\d_0)I,\\
  R'&=\g_0 I,\\
  C'&=\b SI,
\end{align}
where $S(0)=310,000,000$, $C(0)=1896$, $\b=0.64/S(0)$, $\g_0=0.27$, and $\d_0=0.02$.
The equation for $R$ can be decoupled from the system.
The two unknown parameters $q$ and $I(0)$ are estimated via MATLAB's nonlinear least-squares curve fitting solver \texttt{lsqnonlin}. With a random initial guess, the \texttt{lsqnonlin}  solver converges to the (local) optimally estimated parameters: {$q=0.03, I(0)=1070$}.
Numerically, we observe the fitted parameters $q$ and $I(0)$ are insensitive to the chosen initial guess as well as the possible different choices of $R(0)$. It is tempted to treat $R(0)$ as an additional unknown parameter, but in this way its fitted value is not uniquely determined by the given data.
Hence, we reasonably set $R(0)=0$ by assuming nobody  recovered  at the beginning.
Figure \ref{Fitq} illustrates the close match between the reported $C(t)$ data  and simulated $C(t)$  based on the fitted parameters. We point out that the stay-at-home term $qS$ is crucial for achieving such a satisfactory fitting while fixing the other system parameters.
The assumed and estimated parameter values are listed in Table \ref{tab-par}.

\begin{table}
\centering
  \begin{tabular}{|c|c|c|c|}
  	\hline
    Parameter & Symbol & Value & Reference\\
    \hline\hline
    initial susceptible GP & $S_1(0)$ & 310,000,000 & \cite{Population}\\
    initial susceptible HCW & $S_2(0)$ & 16,000,000 & \cite{HCW}\\
    initial infected GP & $I_1(0)$ & 1,896 & \cite{Cases-in-US}\\
    initial infected HCW & $I_2(0)$ & 0 & assumed\\
    initial recovered GP & $R_1(0)$ & 0 & assumed\\
    initial recovered HCW & $R_2(0)$ & 0 & assumed\\
    basic reproduction number & $R_0$ & 2.2 & \cite{Li2020NEJM}\\
    transmission rate & $\b$ & 0.64/S(0) & fitted\\
    death rate & $\d_0/(\g_0+\d_0)$ & 7\% & \cite{DeathRate}\\
    recovery rate & $\g_0$ & 0.27 & fitted\\
    per capital death rate & $\d_0$ & 0.02 & fitted\\
    minimal recovery rate & $\g_\infty$ & 0.1 & assumed\\
    maximal death rate & $\d_\infty$ & 0.1 & assumed\\
    face mask efficacy & $\a$ & 0.9 & assumed\\
    HCW risk factor & $r$ & 3 & assumed\\
    stay-at-home rate & $q$ & 0.03 & fitted\\
    \hline
  \end{tabular}
  \caption{Selection and estimation of model parameters.}
  \label{tab-par}
\end{table}

\section{Numerical Results}
In this section, we report some inspiring simulation results based on our proposed optimal control model for optimizing the allocation of face masks among HCW and GP with our estimated parameters. We highlight that the application of our model to different countries/areas may need to re-estimate some of the parameters based on the actual reported data and population scale. Our tested choices of maximum daily production capacity of the face masks is only for demonstrating our proposed model, which does not reflect the real-life situations. Moreover, our current model does not take into account those home-made cloth face masks and the effects of mandatory quarantine.
Therefore, our model outcomes are mainly for qualitative comparison analysis.

We set $T=100$ and consider 3 scenarios of maximum daily production capacity of face masks:
\begin{itemize}
	\item[(i)] $K_{\max}=16,000,000$ (each HCW can have at most one face mask every day);
	\item[(ii)] $K_{\max}=80,000,000$ (each HCW can have at most five face masks every day);
	\item[(iii)] $K_{\max}=160,000,000$ (each HCW can have at most ten face masks every day).
\end{itemize}

 \begin{figure}[H]
 	\begin{center}
 		\includegraphics[width=0.99\textwidth]{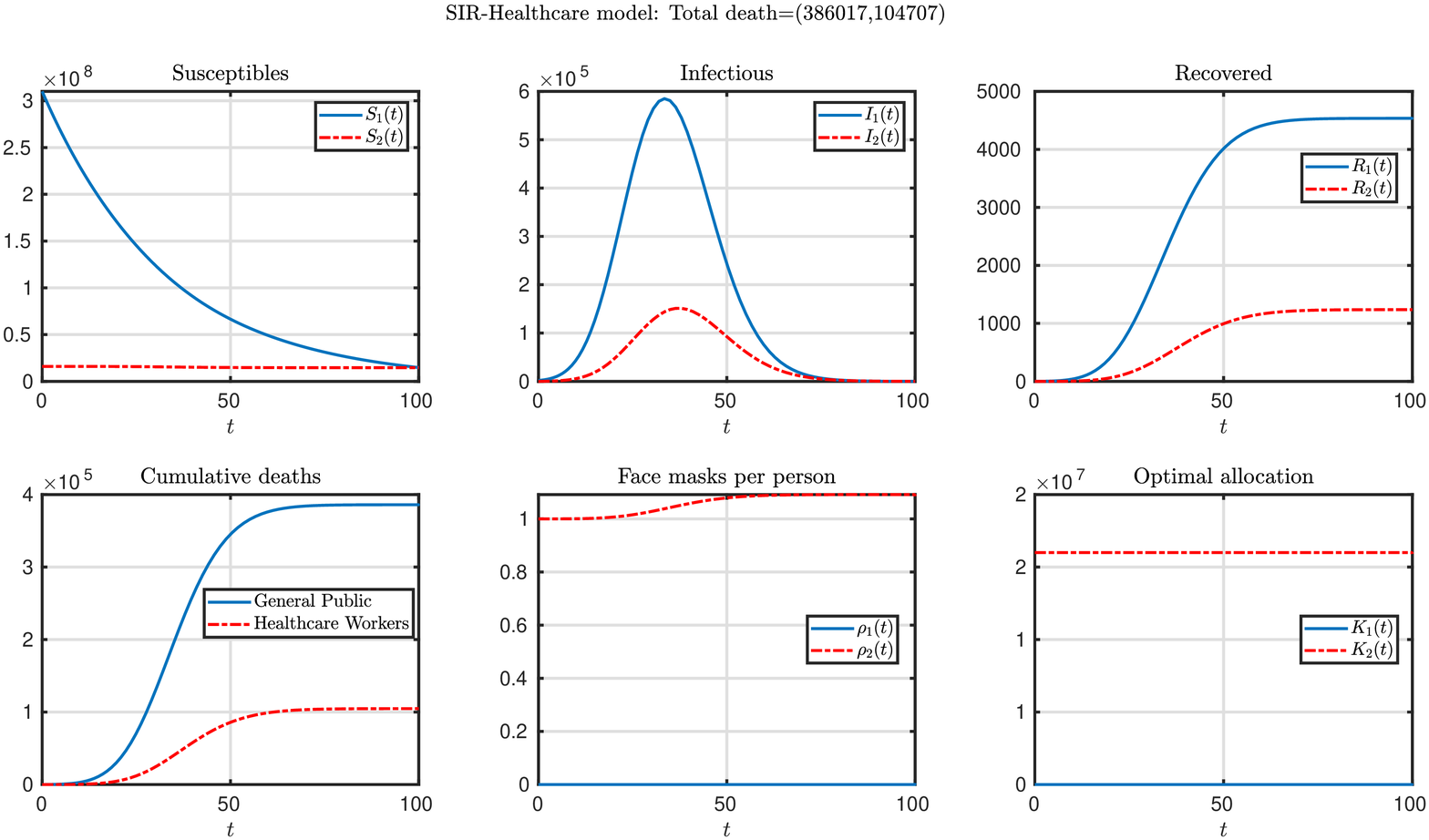}
 	\end{center}
 	\caption{Scenario (i): optimal allocation of face masks between the GP and HCW.}
 	\label{Scenario1}
 \end{figure}

For the scenario (i) with very limit supply, our optimal control results are reported in Figure \ref{Scenario1}. It shows
the face masks should be distributed to the HCW only, which indeed agrees with the guideline from CDC. With this policy control, the total death numbers of the GP and the HCW are $J_1=386,017$ and $J_2=104,707$, respectively.
With more investments from the government and industries, the supply of face masks has been quickly boosted up to a higher level.
In this context, we would want to ask whether this CDC guideline continues to be optimal with increased supply of face masks?
The simple answer is not anymore, as clearly shown in the next two scenarios.

\begin{figure}[H]
	\begin{center}
		\includegraphics[width=0.99\textwidth]{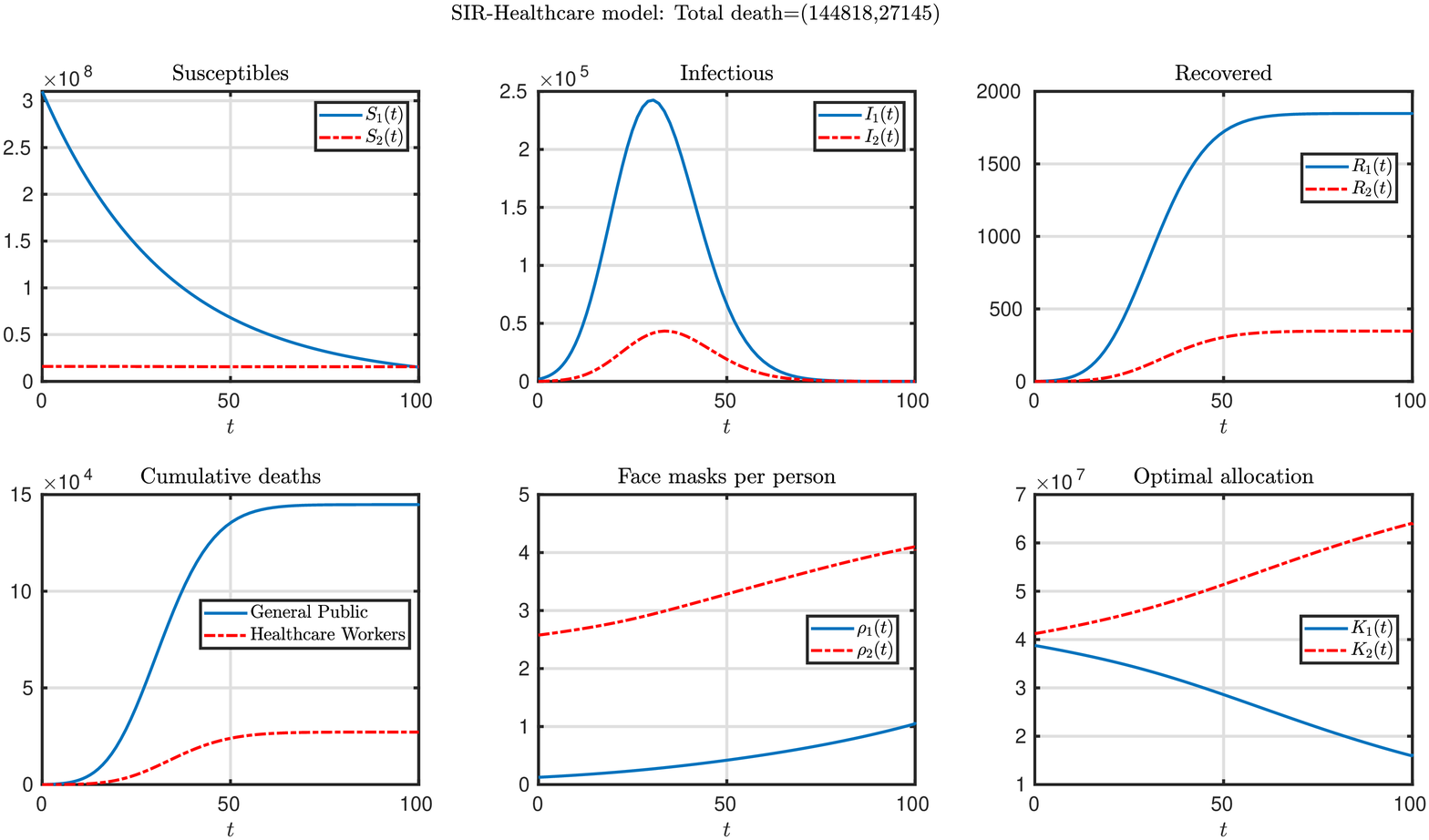}
	\end{center}
	\caption{Scenario (ii)-Optimal: optimal allocation of face masks between the GP and HCW.}
	\label{Scenario2a}
\end{figure}

\begin{figure}[H]
	\begin{center}
		\includegraphics[width=0.99\textwidth]{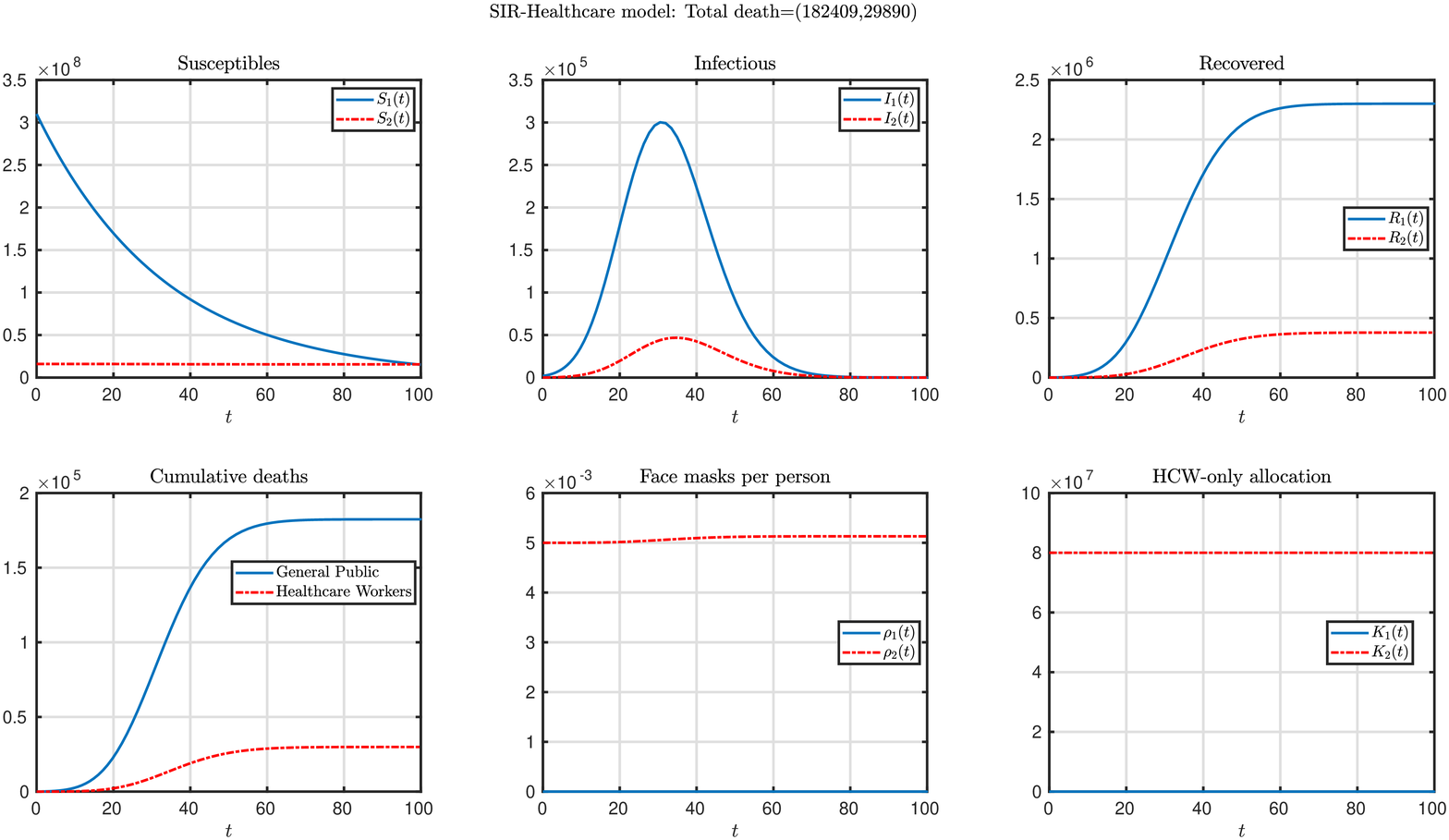}
	\end{center}
	\caption{Scenario (ii)-CDC: all face masks are reserved for HCW only.}
	\label{Scenario2b}
\end{figure}

For the scenario (ii) with moderate supply, our optimal control results are reported in Figure \ref{Scenario2a}. Quite different from the scenario (i), it shows
the face masks should be about equally distributed at the beginning, and then gradually shifted more but not all face masks to the HCW across the whole pandemic outbreak. The total death numbers of the GP and the HCW are $J_1=144,818$ and $J_2=27,145$, respectively.
As a comparison,
Figure \ref{Scenario2b} shows the corresponding outcomes if strictly following the CDC guideline to allocate all face masks to HCW,
where the total death numbers of the GP and the HCW become $J_1=182,409$ (with 26\% increase) and $J_2=29,890$ (with 10\% increase), respectively.
The significant increasing (about 23\%) in the total death toll is somewhat surprising but reasonable, since more infected numbers in GP due to not wearing face masks would infect more HCW and hence lead to higher death rates and more deaths in both groups.


\begin{figure}[H]
	\begin{center}
		\includegraphics[width=0.99\textwidth]{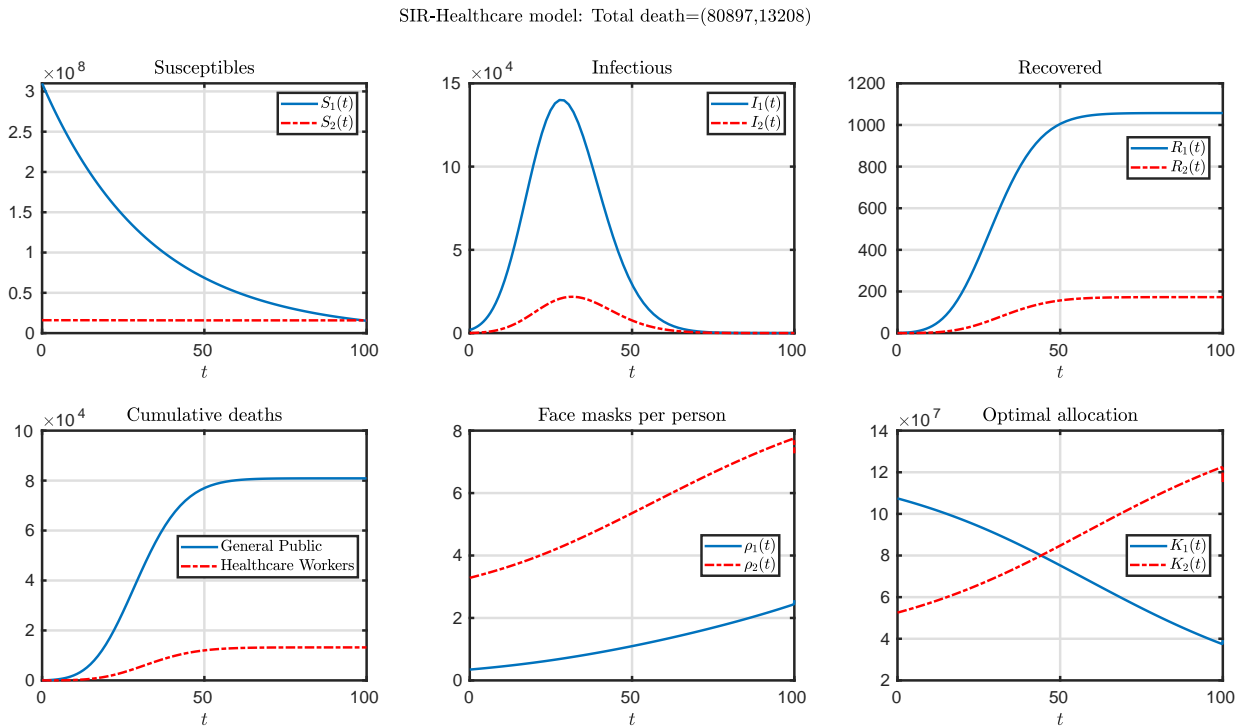}
	\end{center}
	\caption{Scenario (iii)-Optimal: optimal allocation of face masks between the GP and HCW. }
	\label{Scenario3a}
\end{figure}

\begin{figure}[H]
	\begin{center}
		\includegraphics[width=0.99\textwidth]{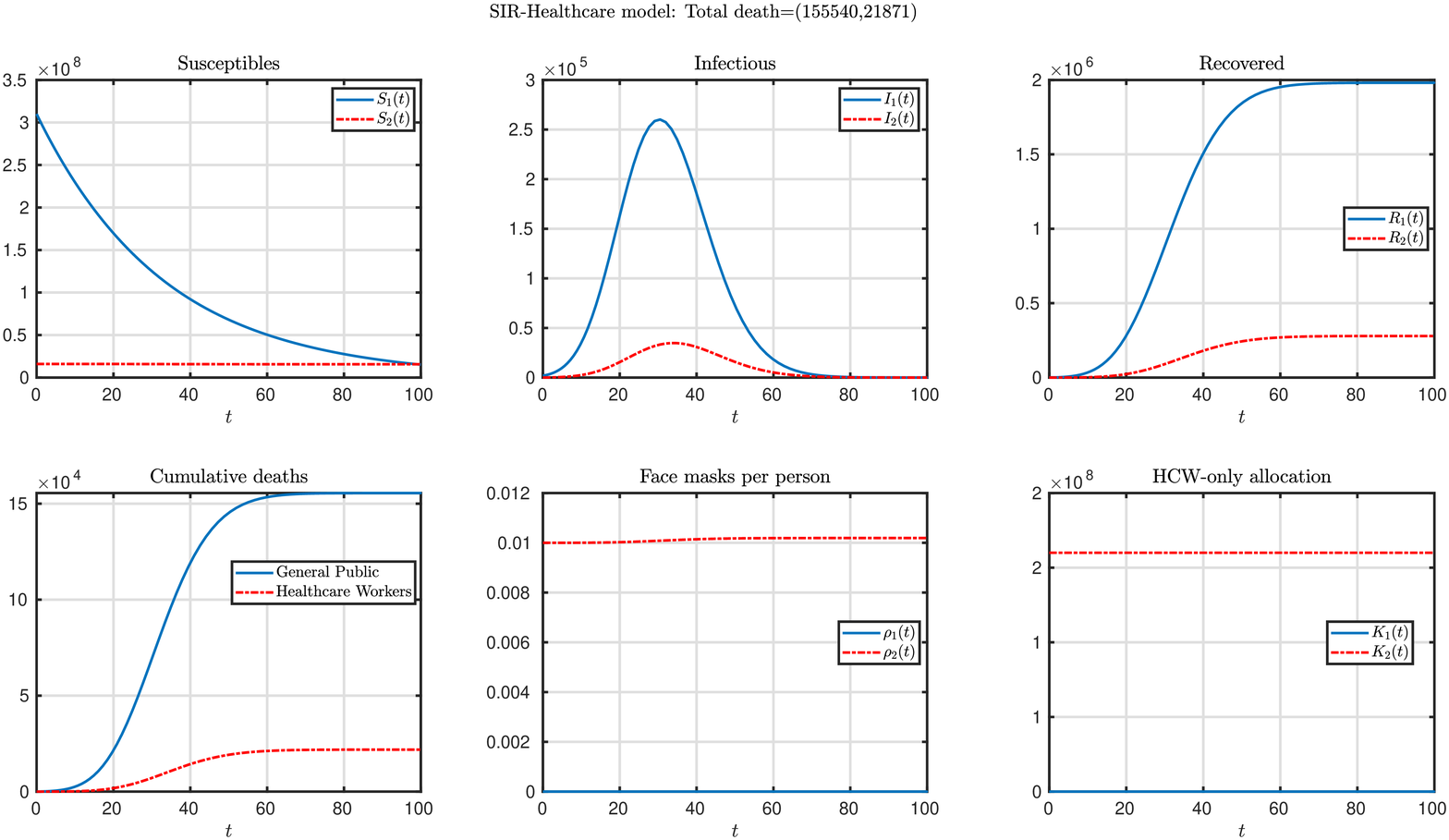}
	\end{center}
	\caption{Scenario (iii)-CDC: all face masks are reserved for HCW only. }
	\label{Scenario3b}
\end{figure}

For the scenario (iii) with very sufficient supply, our optimal control results are reported in Figure \ref{Scenario3a}.
Different from both the scenarios (i) and (ii), it shows
the majority of face masks should be distributed to the GP at the beginning of the epidemic outbreak, and then gradually shifted to the HCW along the outbreak. The total death numbers of the GP and the HCW are $J_1=80,897$ and $J_2=12,208$, respectively. For comparison,
Figure \ref{Scenario3b} shows the corresponding outcomes if following the CDC guideline to allocate all face masks to HCW,
where the total death numbers of the GP and the HCW become $J_1=155,540$ (with 92\% increase) and $J_2=21,871$ (with 80\% increase), respectively.
Astonishingly, the total death numbers are almost doubled if strictly following the reasonably sounding CDC guideline,
where the protection effect of wearing too many face masks for HCW is essentially saturated based on our model setting.
It is also worthwhile to notice that the optimal allocation of masks has greatly flatted the infectious peak curves and hence lead to much less deaths.

\begin{figure}[H]
 	\begin{center}
 		\includegraphics[width=0.90\textwidth]{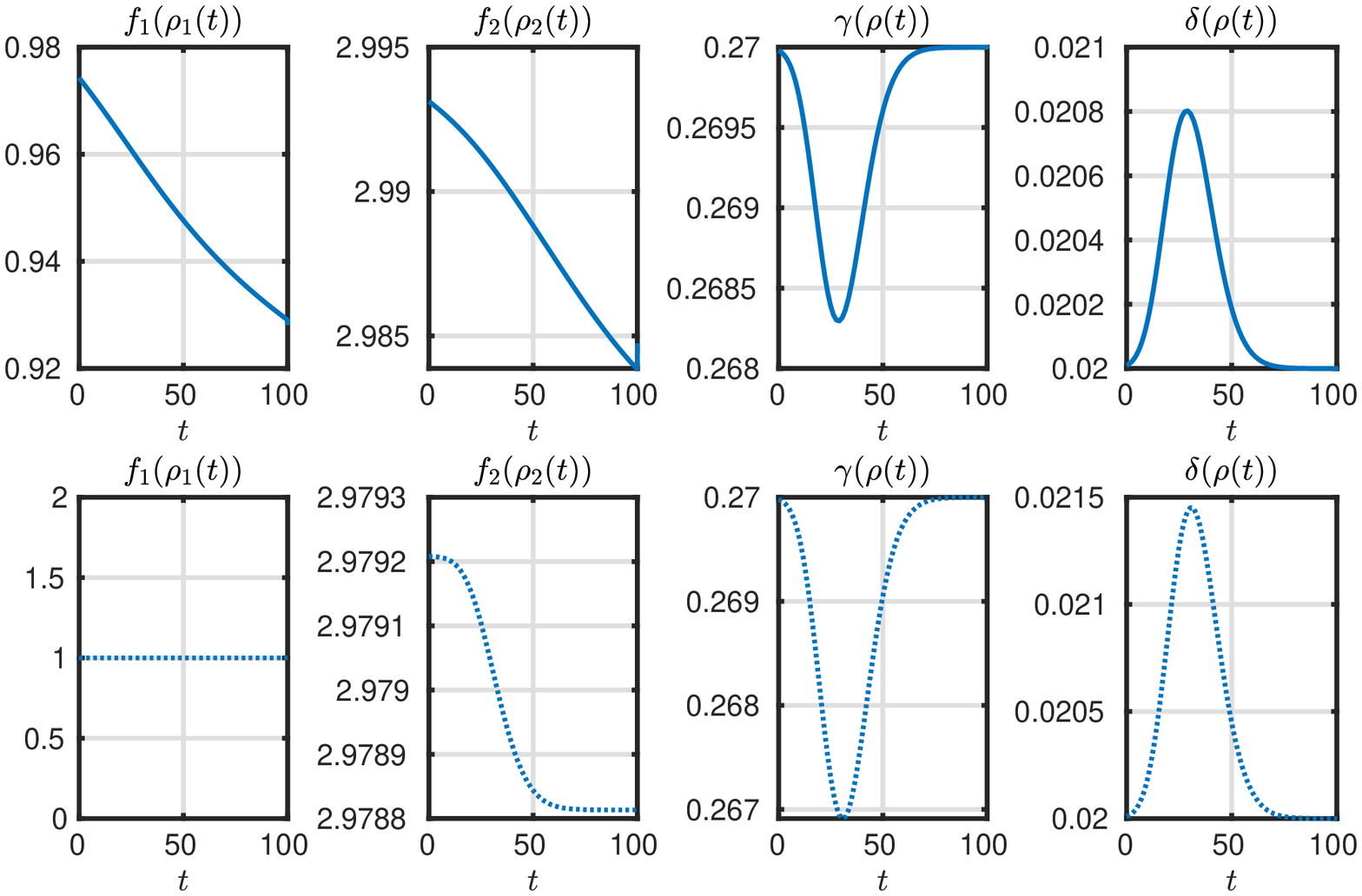}
 	\end{center}
 	\caption{The functions $f_1,f_2,\gamma,\delta$ w.r.t time (Top row: optimal; Bottom row: HCW only). }
 	\label{Ratesfuncs}
 \end{figure}

As an alternative illustration, the relevant functions $f_1,f_2,\gamma,\delta$ with respect to time
are compared in Figure \ref{Ratesfuncs}, where we indeed observe slightly higher death rates and lower recovering rates if all face masks are reserved for HCW only (as advised by CDC).
The scenario (iii) indicates that an appropriately balanced allocation of the face masks between GP and HCW plays a significant role in saving more lives.
In summary, three different scenarios manifest the allocation of face masks
needs to be carefully optimized, especially when the supply becomes gradually more sufficient.

\section{Conclusion and Discussion}
In this paper, we constructed a two-group SIR model to optimize the distribution of face masks among the healthcare workers (HCW) and the general public (GP). When the supply of face masks is in short, our result indicates that all face masks should be reserved for the HCW. This coincides with the advice from the CDC in April 2020  \cite{Protect}. However, when there are plenty supply of face masks, the general public should share a large portion of face masks at the beginning of an epidemic outbreak. This result somewhat contradicts the recommendations given in March 2020 by the US Surgeon General \cite{StopBuying} and the CDC \cite{MaskNoProtection}. The optimality of this reasonable sound CDC guideline highly depends on the supply level of face masks that changes frequently and varies by locations, and hence this guideline should be modified according to the supply of face masks.
Based on our choices and estimations of parameter values and assuming that the supply of face masks is sufficient and the stay-at-home policy remains effective, our model indicates that the first epidemic wave would have ended in May 2020 with a cumulative
total 93,105 deaths, with 80,897 deaths from the GP and 13,208 deaths from the HCW. Note that the stay-at-home policy was released before the end of the first epidemic wave, and a even stronger second epidemic wave arised afterwards. Unlike physical phenomenon which can be observed from repeated experiments, epidemic outbreaks cannot be tested multiply times. Our model analysis and numerical simulation provide theoretical experiments on what would have been occurred if the general public were advised to wear face masks at the beginning of the first epidemic wave and the stay-at-home policy were enforced until the end of the epidemic wave.

There are some limitations for our studies. For instance, the values for parameters $\d_\infty,\g_\infty,\a,r$ are arbitrarily chosen. More real-life data collections are required to obtain more reliable estimations on the death and recovery rates of SARS-CoV-2 in the case of overwhelmed healthcare system, the efficacy of wearing face masks, and the risk of healthcare workers.
With these limitations being said, our model predictions may vary a lot whenever the involved parameters are changed. In particular, the reopening of the country will significantly diminish the stay-at-home efficacy and general multiple epidemic waves.

Our proposed model can be extended in several different ways.
For example, it is interesting to alternatively enforce an integral constraint
$
\int_{0}^t K_1(\tau)+K_2(\tau) d\tau \le (K_{0}+K_{\max}t),
$
which is more practical since it allows flexible usage of face masks according to the epidemic dynamics and possible initially stocked face masks (denoted by $K_0$).
Our developed model can be generalized to optimize the allocations of other personal protective equipments (PPE) among various groups of populations (e.g., public workers vs the general public).
It is also possible to further differentiate the protection effects between different types of face masks.
In addition, the objective functional may also includes some economic considerations, such as the production and transportation costs of face masks.
The generalization of our model to optimally allocate the limited vaccination is also interesting.


\end{document}